\RequirePackage{silence}
\documentclass[twocolumn]{aastex631}
\usepackage{enumitem}
\usepackage{array}
\usepackage{amsmath, amssymb}
\usepackage{longtable}
\usepackage{graphicx}

\setlist[enumerate]{itemsep=0pt,parsep=0pt}
\begin{document}

\title{A Uniform Census of Variable Stars in M101 Based on HST/ACS Photometry}

\author{Jiyu Wang}
\affiliation{CAS Key Laboratory of Optical Astronomy, National Astronomical Observatories, Chinese Academy of Sciences, Beijing 100101, China}
\affiliation{School of Astronomy and Space Science, University of the Chinese Academy of Sciences, Beijing, 100049, China}

\author{Xiaodian Chen}
\affiliation{CAS Key Laboratory of Optical Astronomy, National Astronomical Observatories, Chinese Academy of Sciences, Beijing 100101, China}
\affiliation{School of Astronomy and Space Science, University of the Chinese Academy of Sciences, Beijing, 100049, China}
\affiliation{Institute for Frontiers in Astronomy and Astrophysics, Beijing Normal University, Beijing 102206, China}

\author{Pinjian Chen}
\affiliation{CAS Key Laboratory of Optical Astronomy, National Astronomical Observatories, Chinese Academy of Sciences, Beijing 100101, China}
\affiliation{School of Astronomy and Space Science, University of the Chinese Academy of Sciences, Beijing, 100049, China}

\author{Jianxing Zhang}
\affiliation{CAS Key Laboratory of Optical Astronomy, National Astronomical Observatories, Chinese Academy of Sciences, Beijing 100101, China}
\affiliation{School of Astronomy and Space Science, University of the Chinese Academy of Sciences, Beijing, 100049, China}

\author{Shu Wang}
\affiliation{CAS Key Laboratory of Optical Astronomy, National Astronomical Observatories, Chinese Academy of Sciences, Beijing 100101, China}
\affiliation{School of Astronomy and Space Science, University of the Chinese Academy of Sciences, Beijing, 100049, China}

\author{Licai Deng}
\affiliation{CAS Key Laboratory of Optical Astronomy, National Astronomical Observatories, Chinese Academy of Sciences, Beijing 100101, China}
\affiliation{School of Astronomy and Space Science, University of the Chinese Academy of Sciences, Beijing, 100049, China}
\affiliation{Department of Astronomy, China West Normal University, Nanchong, 637009, China}

\correspondingauthor{Xiaodian Chen}
\email{chenxiaodian@nao.cas.cn}

\begin{abstract}
We present a homogeneous HST/ACS F555W/F814W time-series census of variable stars in two M101 disk fields, using a uniform multi-band reduction and reproducible variable-selection workflow for nearby SN-host galaxies. Supplementary archival F555W epochs are used only to refine selected long-period solutions. Candidates are selected with Lomb--Scargle period searches, Fourier light-curve modeling, two-band consistency statistics, artificial-star-test photometric corrections, and mode-aware Cepheid refinement. The final catalog contains 1417 variable-star candidates, including 1102 secure fundamental-mode and 78 secure first-overtone classical Cepheids. Of these secure Cepheids, 420 are newly identified relative to the Shappee \& Stanek catalog. The catalog also contains luminous candidates outside the classical Cepheid locus. We release the catalog and source-level two-band light curves for studies of Cepheid physics, stellar variability, and distance-scale applications.
\end{abstract}


\section{Introduction}
Variable stars are key time-domain tracers for studying stellar evolution, stellar populations, and galactic structure. Measurements of their periods, amplitudes, and mean magnitudes provide direct constraints on internal structure, pulsation mechanisms, and evolutionary state. Moreover, several classes of variables obey empirical relations that enable distance calibration and thus play a central role in establishing the local distance scale \citep{1912HarCi.173....1L,Freedman_2001,2016ApJ...826...56R,2026RAA....26h4003C}.

In nearby galaxies beyond the Local Group, variability studies have historically centered on classical Cepheids because their period--luminosity relation is a cornerstone of the extragalactic distance scale. As a result, many earlier searches were optimized for Cepheid discovery and distance work, rather than for a homogeneous census of the broader luminous-variable population.

Space-based time-domain observations and improved crowded-field photometry now make such censuses more practical. The angular resolution and photometric stability of the Hubble Space Telescope (HST), together with advances in PSF-fitting photometry, period-search pipelines, artificial-star-test-based bias correction, and sample-refinement workflows, enable more uniform reanalysis of archival data. Improved detection efficiency and workflow reproducibility make it timely to revisit well-observed nearby galaxies under consistent criteria \citep{2020ApJS..249...18C,2025ApJS..279...56W}.

M101 (NGC~5457) is a nearby, nearly face-on grand-design spiral galaxy and a benchmark system for resolved-stellar studies. It has long been central to distance-scale work: it was included in the HST Key Project \citep{1996ApJ...463...26K,1998ApJ...508..491S} and later revisited with HST/ACS time-series data by \citet{2011ApJ...733..124S}, who published a widely used catalog of 827 Cepheids in two disk fields. M101 also provides a useful setting for studies of disk structure and evolution, including its radial abundance gradient \citep{2003ApJ...591..801K} and interaction-related outer-disk asymmetry \citep{1997ApJ...481..169W}. Its role in distance-scale work was reinforced by the nearby Type~Ia supernova SN~2011fe, which links resolved-stellar indicators to SN~Ia distances \citep{2012ApJ...760L..14L,2019ApJ...885..141B}.

This paper is organized as follows. Section 2 describes the observations and photometry. Section 3 defines the methodological framework and quantitative checks for the variable-star search. Section 4 presents the implemented selection criteria and period determination. Section 5 outlines sample refinement, secure Cepheid selection, candidate-level labeling of non-Cepheid variables, and the released catalog. Section 6 provides a discussion, including a comparison with previous Cepheid studies. Section 7 presents the conclusions.

\begin{deluxetable*}{llllcccll}
\tabletypesize{\scriptsize}
\tablecaption{Summary of \textit{HST}/ACS observations used in this work.\label{tab:obslog}}
\tablehead{
\colhead{Field} &
\colhead{Program} &
\colhead{PI} &
\colhead{Instrument/Filter(s)} &
\colhead{$N_{\rm epoch}$} &
\colhead{$N_{\rm exp,used}$} &
\colhead{$t_{\rm exp}$ (s)} &
\colhead{Date(s)} &
\colhead{Notes}
}
\startdata
F1 & 10918 & Freedman & ACS/WFC, F555W+F814W & 12 & 24 & 665/362 & 2006-12-23--2007-01-21 & Primary two-band time series. \\
F1 & 9490  & Kuntz    & ACS/WFC, F555W       & 2  & 4  & 360     & 2002-11-15, 2002-11-16 & Adjacent pointings (mosaic)\tablenotemark{a,c}. \\
F1 & 12880 & Riess    & ACS/WFC, F555W       & 2  & 4  & 375     & 2013-03-11, 2013-03-18 & Supplementary epochs\tablenotemark{b,d}. \\
F1 & 13737 & Shappee  & ACS/WFC, F555W       & 1  & 2  & 446     & 2015-04-06             & Supplementary epoch\tablenotemark{b,e}. \\
\hline
F2 & 10918 & Freedman & ACS/WFC, F555W+F814W & 12 & 24 & 665/362 & 2006-12-25--2007-01-23 & Primary two-band time series. \\
F2 & 9490  & Kuntz    & ACS/WFC, F555W       & 2  & 4  & 360     & 2002-11-13             & Adjacent pointings (mosaic)\tablenotemark{a,c}. \\
F2 & 9492  & Bresolin & ACS/WFC, F555W       & 1  & 2  & 360     & 2003-01-16             & Supplementary epoch\tablenotemark{b}. \\
F2 & 17312 & Huang    & ACS/WFC, F555W       & 3  & 6  & 492--550& 2023-07-19, 2024-08-13, 2024-08-16 & Supplementary epochs\tablenotemark{b}. \\
\enddata
\tablenotetext{a}{Program 9490 consists of adjacent pointings. Multiple pointings are combined to cover the 10918 footprints. Only the portions overlapping the primary fields are used.}
\tablenotetext{b}{Supplementary data are used only to extend the temporal baseline for candidates with $P \gtrsim 25$~days. Only F555W images are used, and we retain two exposures per epoch.}
\tablenotetext{c}{Program 9490 data were previously used by \citet{2005ApJ...620L..31K}.}
\tablenotetext{d}{Program 12880 data were previously used by \citet{2016ApJ...826...56R}.}
\tablenotetext{e}{Program 13737 data were previously reported by \citet{2015ATel.7392....1S} and \citet{2017ApJ...841...48S}.}
\end{deluxetable*}

\section{Observations and Data Reduction}
\subsection{Observations}
\label{sec:obs}

The primary data set used in this work is the \textit{HST} Advanced Camera for Surveys (ACS) time-series program 10918, which targets two disk fields in M101 (hereafter F1 and F2). This program provides multi-epoch imaging in the F555W and F814W bands with a baseline of $\sim$29~days, and it defines the uniform core data set for constructing the parent sample and for the primary variability analysis in this paper. Each primary epoch consists of two consecutive exposures, and the observing cadence and sampling strategy are similar for the two fields.

\begin{figure*}
    \centering
    \includegraphics[width=\linewidth]{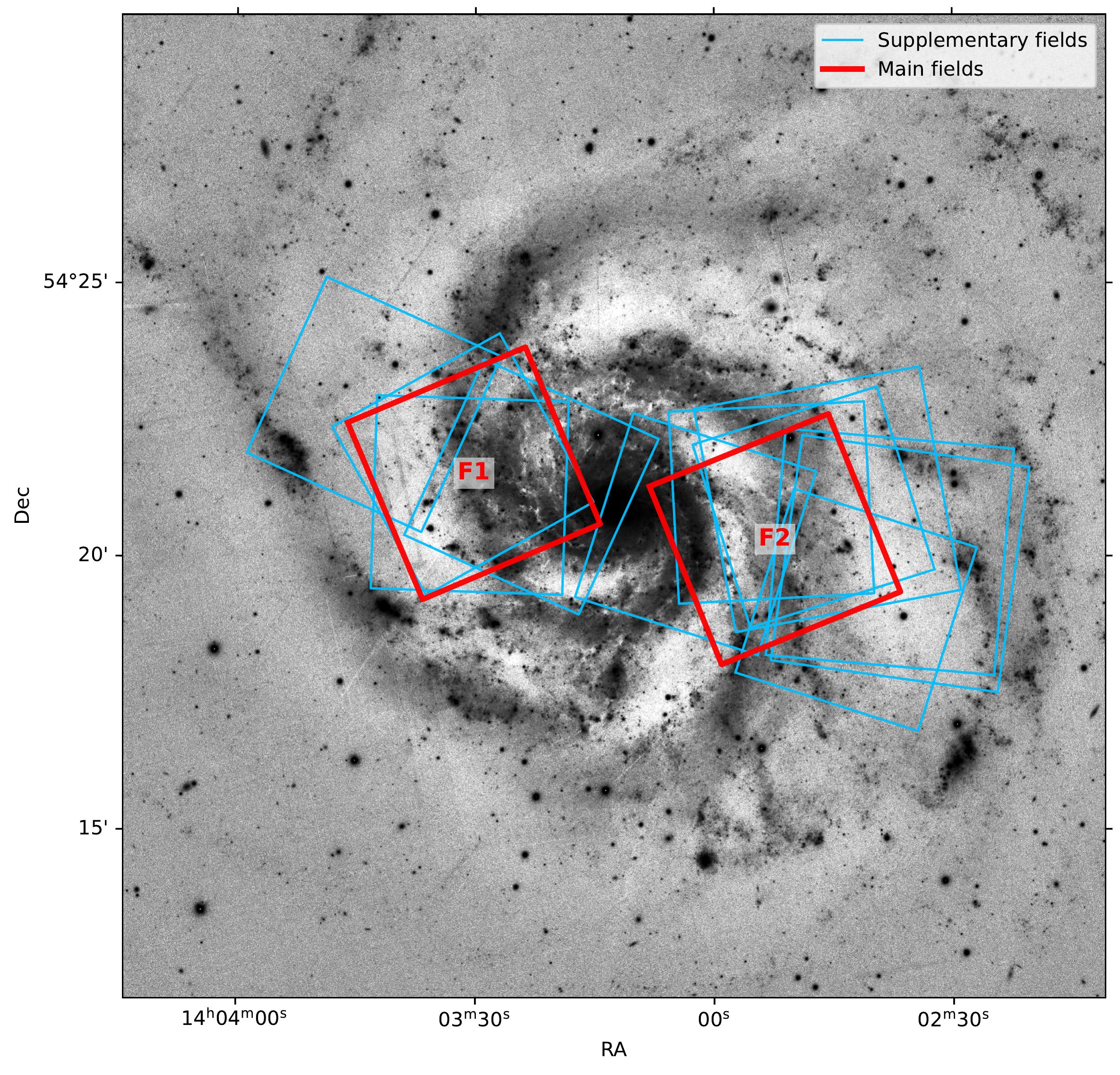}
    \caption{Sky coverage of the HST observations in M101 overlaid on a Pan-STARRS1 background image. The two primary fields are labeled as F1 and F2 (thick red outlines), while supplementary fields are shown with thin blue outlines.}
    \label{fig:m101_footprints}
\end{figure*}

We additionally include supplementary ACS observations from several archival programs. To keep the supplementary measurements consistent with the primary two-exposure-per-epoch structure, we retain two F555W exposures per supplementary epoch. When the available exposures have different integration times, we select the two longest exposures. When the exposure times are the same, we select two consecutive exposures. Some early supplementary programs consist of adjacent pointings, and we use only the portions that overlap the primary fields. Table~\ref{tab:obslog} summarizes the primary and supplementary observations, including program IDs, date ranges, filters, and the adopted number of epochs/exposures. Figure~\ref{fig:m101_footprints} shows the sky coverage of the primary fields and supplementary observations.

The supplementary data are used only as long-baseline F555W anchors for candidate variables with $P \gtrsim 25$~days, whose periods approach the $\sim$29-day span of the primary time series. They are excluded for shorter-period candidates, because the primary data already cover multiple cycles and sparse long-baseline points can introduce additional alias structure when the primary-data period precision is insufficient.

\subsection{Data Reduction and Photometry}
\label{sec:reduction}

All \textit{HST}/ACS data used in this work were retrieved from the Mikulski Archive for Space Telescopes (MAST). The primary HST observations analyzed in this work can be accessed via \dataset[doi:10.17909/wntn-v633]{https://doi.org/10.17909/wntn-v633}, and the supplementary HST observations can be accessed via \dataset[doi:10.17909/jrwf-xh95]{https://doi.org/10.17909/jrwf-xh95}. As inputs for photometry, we use the \texttt{flc.fits} products, i.e., the calibrated, flat-fielded, and charge-transfer-efficiency (CTE) corrected single-exposure images. We perform point-spread-function (PSF) fitting photometry with DOLPHOT (the ACS module). For each M101 field, the primary F555W/F814W exposures and the supplementary F555W exposures are tied to the same astrometric reference frame and source list, so that every measurement can be tracked with a common source identifier (\texttt{global\_id}). This source-level organization places the primary and supplementary measurements on the same DOLPHOT photometric system and avoids constructing an independent supplementary catalog that would require a separate cross-match. Because the two M101 fields (F1 and F2) do not overlap, we process them independently.

The primary program 10918 data define the uniform parent sample and are used for the main two-band variability analysis. The supplementary measurements are attached only to sources already identified in the primary catalog and are used only when re-examining candidates with $P \gtrsim 25$~days. Thus, supplementary data can change the adopted long-period solution when the extended baseline yields a stable, localized periodogram peak and a credible folded F555W light curve, but they do not introduce new sources into the parent sample and are not used in the two-band sample refinement or final candidate labeling.

Prior to photometry, we pre-align all input images using \texttt{TweakReg} to reduce pointing offsets between visits and epochs and to stabilize subsequent processing. We then construct a high signal-to-noise reference frame from the primary data: we select six F555W \texttt{flc.fits} exposures from the primary field, drizzle-combine the \emph{aligned} images with \texttt{AstroDrizzle}, and generate a deep \texttt{drc.fits} reference image with a total exposure time exceeding 3000~s. DOLPHOT uses this reference frame to define the astrometric coordinate system and to perform global PSF fitting across epochs. The resulting time-series table contains primary F555W/F814W measurements for the full variability search and, for the subset of long-period candidates, the supplementary F555W measurements associated with the same \texttt{global\_id}.

\section{Methodological Framework and Quantitative Checks}

A variable-star census in a crowded SN-host galaxy is limited by several linked methodological choices. To reduce dependence on open-ended visual judgment, we organize the search around four quantifiable requirements: (1) source-finding completeness in the input photometric catalog, (2) completeness and reliability of variable selection, (3) accuracy of Cepheid refinement and variable-class assignment, and (4) reproducibility of the full workflow through explicit thresholds and released diagnostics.

The first requirement is source-finding completeness. A variable star cannot be recovered if it never enters the photometric catalog or lacks enough valid epochs for a light curve. We address this by using a multi-epoch DOLPHOT catalog tied to a common astrometric frame and source identifier, rather than a single-epoch reference-catalog entrance. In such HST crowded-field reductions, a deep reference image is commonly built from multiple aligned exposures and then used to define the astrometric system and source list for PSF photometry on the individual science images \citep[e.g.,][]{2000PASP..112.1383D,2016ascl.soft08013D,2014ApJS..215....9W,2023MNRAS.521.1532J}. This carries source positions and neighbor constraints through the time series, helping retain faint or blended sources and variables that may be near minimum light in any single reference epoch. The corresponding quantitative checks are the number of detected sources before quality cuts, the comparison with a single-epoch entrance catalog, and the AST completeness curves in \texttt{F555W} and \texttt{F814W}. In Field~F1, the global multi-epoch catalog contains 1,050,915 detected sources before quality cuts, whereas a representative two-band single-epoch DOLPHOT run under the same basic detection definition yields 643,206 sources. Thus, a two-band single-epoch reference-catalog entrance would miss 407,709 sources, or 38.8\% of the multi-epoch source list, before any variability information is used. Such source-finding loss can propagate directly into the subsequent variable-star search.

The second requirement is the completeness and reliability of variable selection. A selection based only on a variability index or broad light-curve scatter can miss many faint periodic variables whose amplitudes are comparable to the photometric noise, whereas a selection that combines scatter, period-search significance, light-curve model preference, and multi-band coherence can improve completeness while controlling false positives \citep[e.g.,][]{1996PASP..108..851S,2020ApJS..249...18C,2025ApJS..276...57G,2025ApJS..279...56W}. We therefore separate a completeness-oriented parent-sample and candidate-generation step from the subsequent automated confirmation tests. The parent-sample quality cuts follow the completeness-oriented DOLPHOT-guide criteria in sharpness, crowding, flag, and object type. The deliberate relaxation is the use of ${\rm S/N}\ge2$ instead of ${\rm S/N}\ge4$, which extends the search to sources near \texttt{F555W}$\sim28$~mag. The quantitative selection diagnostics are the pre-selection counts in the \texttt{F555W} magnitude--scatter plane, the number of candidates surviving the \texttt{F555W} period-search and Fourier-fit criteria, the number surviving two-band coherence tests, and the number removed by final visual vetting. The final visual step is therefore limited to cases with poor folded light-curve quality, poorly constrained Fourier morphology, or residual two-band inconsistency, and its effect is explicitly counted rather than left as an open-ended judgment. We evaluate the sensitivity of the final secure Cepheid sample to stricter two-band S/N and crowding cuts in Section~\ref{sec:comparison_previous}.

We also quantify the false-positive behavior of the variable-selection stage with an end-to-end pure-noise simulation. We simulate non-variable sources using the actual cadence, filter coverage, and photometric uncertainties of real light curves across the observed \texttt{F555W} magnitude range. Each measurement is replaced by a constant mean plus Gaussian noise with $\sigma_{\rm sim}=(\sigma_{\rm mag}^2+0.02^2)^{1/2}$, and the simulated light curves are passed through the same \texttt{F555W} LS search, Fourier-model selection, and two-band consistency criteria used below. In 14,810 pure-noise realizations, only 22 cases pass the \texttt{F555W}-only period-search and Fourier-fit criteria (0.149\%), and none pass the subsequent two-band consistency requirement. This test does not require the individual cuts to be statistically independent. Instead, it evaluates the combined automated selection on noise-dominated light curves.

The third requirement is classification accuracy. If all periodic candidates are treated as a single Cepheid-like sample, the number of recoverable variables can be artificially limited and the secure Cepheid sequence can be contaminated by other variable types or by Cepheids with different pulsation modes. We therefore treat variable detection, F/1O mode assignment, and final Cepheid refinement as separate steps. The measurable diagnostics are the AST-corrected color and Wesenheit quantities, the held-out performance of the F/1O random-forest classifier, the separation of the final F and 1O samples in Fourier-parameter and PW space, and the final 3$\sigma$ PW refinement after mode correction. In particular, the random-forest classifier is trained on OGLE LMC Cepheids and reaches a held-out balanced accuracy of 0.9931. When applying it to M101, the F-mode reference ridge uses a fixed slope and an intercept determined from the high-purity subset cross-matched to \citet{2011ApJ...733..124S}, so the classifier uses a relative M101 PW residual rather than an external distance modulus. Sources outside the secure PW sequence are retained in the broader variable catalog and labeled only at the candidate-population level.

The fourth requirement is reproducibility. To minimize dependence on subjective inspection, the catalog construction is based on explicit source-level identifiers, quality cuts, sampling requirements, LS/Fourier model-selection statistics, two-band coherence statistics, AST corrections, RF features, and PW-refinement criteria. The visual-vetting stage is retained only as a final guard against failure modes not fully captured by scalar statistics, and its numerical impact is reported. We release the summary catalog and source-level two-band light curves, including the principal diagnostics used in the selection and classification procedure, so that the period solutions, two-band consistency, Cepheid refinement, and non-Cepheid candidate labels can be re-examined directly.

\section{Variable-star Selection and Period Determination}

\subsection{Parent sample and \texttt{F555W} primary selection}
\label{subsec:f555w_primary_selection}

Following the design above, for each measurement, we require, in sequence:
\begin{enumerate}
\item ${\rm S/N} \ge 2$,
\item ${\rm Sharpness}^2 \le 0.1$,
\item ${\rm Crowding} \le 2.25$,
\item photometry quality flag $\le 3$,
\item object type $\le 2$,
\item ${\rm Roundness} \le 2$.
\end{enumerate}
After these basic two-band quality cuts, 964,071 sources in Field~F1 and 975,203 in Field~F2 remain for variability screening. The multi-epoch measurements are organized by the global source identifier (\texttt{global\_id}) defined in the photometric catalog.

\begin{figure*}[t]
    \centering
    \includegraphics[width=0.95\textwidth]{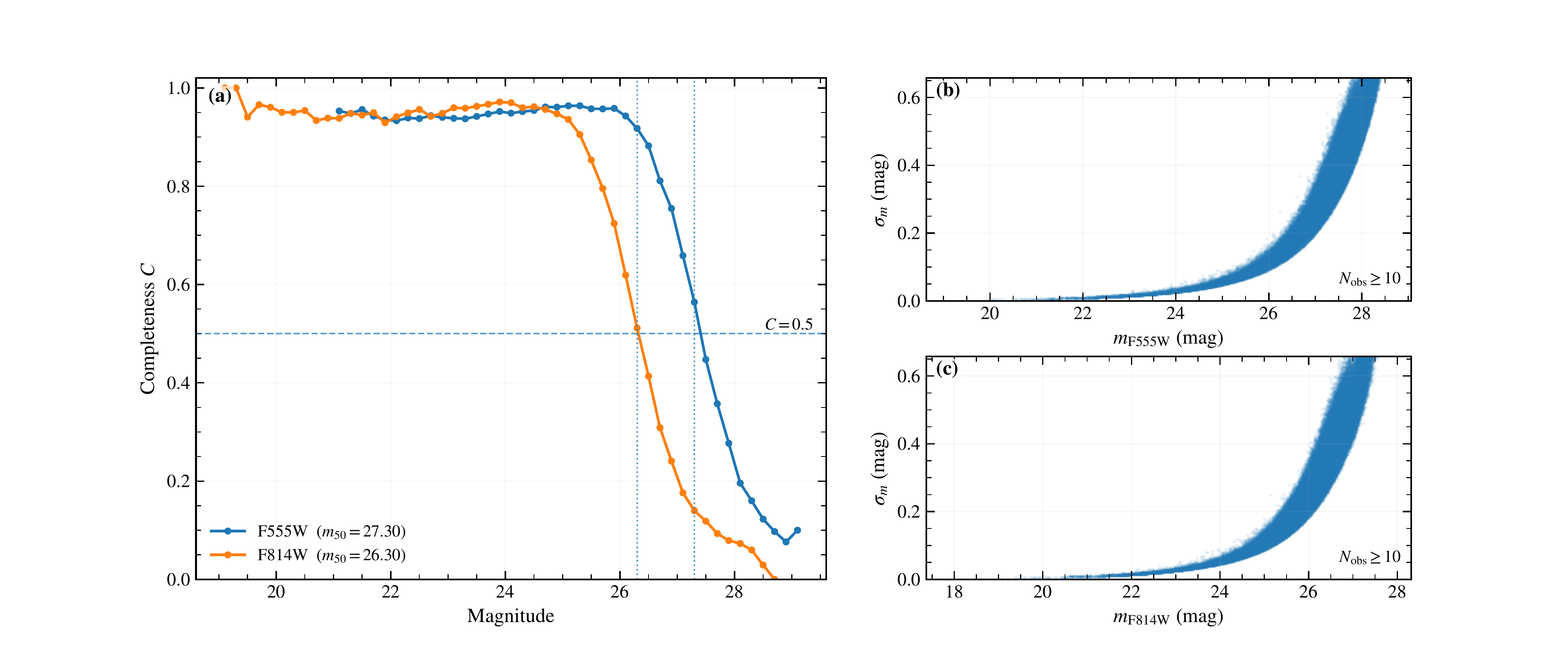}
    \caption{Artificial-star test results for the M101 fields. Panel (a) shows the completeness $C$ as a function of input magnitude in \texttt{F555W} and \texttt{F814W}, with $C=0.5$ and the corresponding $m_{50}$ values indicated. Panels (b--c) show the magnitude--uncertainty relation ($\sigma_m$) for sources with sufficient sampling.}
    \label{fig:ast_m101f2}
\end{figure*}

Before selecting variables, we run field-wide artificial-star tests (ASTs) in both fields by injecting artificial sources across the full footprint and reprocessing them with the same photometric pipeline \citep{2014ApJS..215....9W,2018ApJ...852...60J}. The ASTs provide completeness as a function of input magnitude and the magnitude--uncertainty relation. Figure~\ref{fig:ast_m101f2} shows the completeness curves in \texttt{F555W}/\texttt{F814W} and the corresponding uncertainty distributions. To ensure adequate temporal sampling, we require
\begin{align}
N_{\mathrm{F555W}} &\ge 15, \\
N_{\mathrm{epoch}} &\ge 8, \\
N_{\mathrm{F814W}} &\ge 12,
\end{align}
where $N_{\mathrm{F555W}}$ and $N_{\mathrm{F814W}}$ are the numbers of valid measurements passing the above cuts (out of 24 visits), and $N_{\mathrm{epoch}}$ is the number of distinct epochs.

For each source, let $m$ denote the \texttt{F555W} time-series mean magnitude, $\sigma_m$ the typical formal photometric uncertainty (catalog \texttt{mag\_err}), and $s_m \equiv \mathrm{std}(m_i)$ the observed scatter (catalog \texttt{mag\_std}). We pre-select variability candidates in the \texttt{F555W} $m$--$s_m$ plane using a sliding-bin method: magnitude bins of width $\Delta m = 0.3$~mag, stepped by 0.1~mag, and selection of the highest-scatter 40\% in each bin. We additionally include sources with $s_m > 3\,\sigma_m$, so that objects with scatter significantly above the expected photometric noise are retained. This deliberately broad pre-selection yields 110,540 sources in Field~F1 and 115,904 in Field~F2. The broad scatter cut is used to avoid rejecting real low-amplitude or faint variables prematurely. Reliability is imposed by the period-search, model-selection, two-band consistency, and visual-vetting steps described below.

Before the period search, we iteratively reject \texttt{F555W} measurements more than $3.5\sigma$ from the median, recalculating the sample standard deviation for at most three iterations. For each candidate, we then run an unweighted Lomb--Scargle (LS) search in \texttt{F555W} and keep the three highest-power local maxima as candidate periods $\{P_1,P_2,P_3\}$. For each $P$, we fit a second-order Fourier model,
\begin{equation}
m(t) = m_0 + \sum_{k=1}^{2}\left[a_k \cos\!\left(\frac{2\pi k t}{P}\right) + b_k \sin\!\left(\frac{2\pi k t}{P}\right)\right],
\end{equation}
along with two non-variable baselines: a constant model $m(t)=\mu_0$ and a linear trend
\begin{equation}
m(t)=\alpha+\beta\,(t-\bar t),
\end{equation}
where $\bar t$ is the mean observation time. Model preference is quantified with the Bayesian Information Criterion (BIC) from unweighted residual sums of squares,
\begin{equation}
\begin{aligned}
\mathrm{BIC}(M) &= n\ln\!\left(\frac{\mathrm{RSS}(M)}{n}\right) + k_M\,\ln n, \\
\mathrm{RSS}(M) &= \sum_{i=1}^{n}\left(m_i-\hat m_M(t_i)\right)^2.
\end{aligned}
\end{equation}
Here $n$ is the number of data points and $k_M$ is the number of free parameters. We compute $\mathrm{BIC}_{\mathrm{const}}$ ($k_{\mathrm{const}}=1$), $\mathrm{BIC}_{\mathrm{lin}}$ ($k_{\mathrm{lin}}=2$), and $\mathrm{BIC}_{\mathrm{model}}$ for the periodic fit. For the fixed-$P$ Fourier model, $k_{\mathrm{model}}=5$ ($m_0,a_1,b_1,a_2,b_2$). We define
\begin{equation}
\mathrm{BIC}_{\mathrm{base}}=\min\!\left(\mathrm{BIC}_{\mathrm{const}},\,\mathrm{BIC}_{\mathrm{lin}}\right),
\end{equation}
\begin{equation}
\Delta\mathrm{BIC}_{\mathrm{best}} \equiv \mathrm{BIC}_{\mathrm{base}}-\mathrm{BIC}_{\mathrm{model}}.
\end{equation}
The period with the largest $\Delta\mathrm{BIC}_{\mathrm{best}}$ is adopted.

We then require the \texttt{F555W}-based solution to satisfy
\begin{enumerate}
\item ${\rm FAP} \le 0.01$,
\item $\Delta\mathrm{BIC}_{\mathrm{best}} \ge 10$,
\item $A_{\mathrm{tot}} \ge 3\,\mathrm{RMSE}$,
\item $R^2 \ge 0.6$.
\end{enumerate}
Here ${\rm FAP}$ is the LS false-alarm probability at the candidate peak, $\mathrm{RMSE}$ is the Fourier-fit residual root-mean-square, and $R^2$ is the coefficient of determination. The amplitude proxy is
\begin{equation}
A_{\mathrm{tot}} \equiv 2\sqrt{a_1^2+b_1^2+a_2^2+b_2^2}.
\end{equation}
Because ${\rm FAP}$ calibration can be imperfect for sparse and heterogeneous sampling, we use the ${\rm FAP}$ cut as a permissive lower bound rather than a standalone discriminator.
After applying these \texttt{F555W}-based period-search and Fourier-fit criteria, 1075 candidates remain in Field~F1 and 1092 candidates remain in Field~F2. These sources are then passed to the two-band consistency test described below.

\subsection{Two-band consistency and final vetting}
\label{subsec:two_band_consistency}

We apply the two-band consistency test only after a source satisfies the \texttt{F555W}-based variability criteria described above. As motivated in Section~3, for each source passing the \texttt{F555W} selection, we construct three two-band coherence statistics, $J_{\rm time}$, $J_{\phi}$, and $f_{\rm same}$. The time-domain statistic tests whether nearly simultaneous \texttt{F555W}/\texttt{F814W} changes are consistent with real variability rather than single-band noise, while the phase-domain statistic tests whether the adopted \texttt{F555W} period also produces a coherent folded \texttt{F814W} response.

Before computing these statistics, we group the cleaned measurements into epoch-level points. Epochs are inferred from the observing times, with a new epoch assigned when the gap between adjacent measurements exceeds 0.5~d. Within each epoch and band, repeated exposures are combined using inverse-variance weighted mean magnitudes. The epoch uncertainty is taken as the larger of the formal weighted error and the within-epoch scatter divided by the square root of the number of exposures. The \texttt{F555W} and \texttt{F814W} epoch-level points are then paired by the same epoch identifier.

For each band $b\in\{555,814\}$, we subtract the band-specific mean magnitude and normalize each epoch-level point by its photometric uncertainty:
\begin{equation}
    \delta_{b,k}
    =
    \frac{m_{b,k}-\bar{m}_{b}}{\sigma_{b,k}},
\end{equation}
where $m_{b,k}$ and $\sigma_{b,k}$ are the epoch-level magnitude and uncertainty for pair $k$, and $\bar{m}_{b}$ is the weighted mean magnitude of that band after the initial cleaning. Since both bands are expressed in magnitudes, the same sign convention is used for \texttt{F555W} and \texttt{F814W}. A positive product of the two normalized residuals therefore indicates that the two measurements are on the same side of their respective band-specific mean magnitudes.

We first define a time-domain consistency statistic. For each matched two-band epoch pair $k$, consisting of one epoch-level \texttt{F555W} point and one epoch-level \texttt{F814W} point, we compute
\begin{equation}
    P_k^{\rm time}
    =
    \delta_{555,k}\,\delta_{814,k} .
\end{equation}
The time-domain coherence statistic is then defined in the same spirit as the Stetson variability index,
\begin{equation}
    J_{\rm time}
    =
    \frac{1}{N_{\rm time}}
    \sum_{k=1}^{N_{\rm time}}
    {\rm sgn}\!\left(P_k^{\rm time}\right)
    \sqrt{\left|P_k^{\rm time}\right|},
\end{equation}
where $N_{\rm time}$ is the number of valid matched epoch pairs. Positive values of $J_{\rm time}$ indicate that the two bands tend to deviate from their mean magnitudes in the same direction at nearly contemporaneous epochs, while values close to zero indicate weak or no two-band coherence.

We also define an analogous phase-domain statistic. The matched epoch pairs are folded with the best period determined from the \texttt{F555W} light curve,
\begin{equation}
    \phi_{k}
    =
    \left[
    \frac{t_{{\rm pair},k}-t_0}{P_{\rm F555W}}
    \right] \bmod 1 ,
\end{equation}
where $t_{{\rm pair},k}$ is the average of the \texttt{F555W} and \texttt{F814W} epoch-level times in pair $k$, and $t_0$ is an arbitrary reference epoch. The folded pairs are divided into phase bins according to the number of available epoch pairs: we use five bins when $N_{\rm pair}\ge10$, four bins when $8\le N_{\rm pair}<10$, and three bins otherwise, requiring at least five pairs to compute $J_{\phi}$. For each phase bin $q$, we compute weighted mean \texttt{F555W} and \texttt{F814W} magnitudes and form normalized residuals $z_{555,q}$ and $z_{814,q}$ relative to the band-specific weighted means. We then define
\begin{equation}
    Q_q
    =
    z_{555,q}\,z_{814,q},
\end{equation}
and compute
\begin{equation}
    J_{\phi}
    =
    \frac{1}{N_{\phi}}
    \sum_{q=1}^{N_{\phi}}
    {\rm sgn}\!\left(Q_q\right)
    \sqrt{\left|Q_q\right|},
\end{equation}
where $N_{\phi}$ is the number of valid phase bins. Compared with $J_{\rm time}$, $J_{\phi}$ is less sensitive to the exact within-epoch time offsets between the two filters, but it relies on the adopted \texttt{F555W} period being approximately correct.

Finally, we define a simpler directional-consistency fraction,
\begin{equation}
    f_{\rm same}
    =
    \frac{1}{N_{\rm pair}}
    \sum_{k=1}^{N_{\rm pair}}
    I\left[
    \delta_{555,k}\,\delta_{814,k} > 0
    \right],
\end{equation}
where $I[\cdot]$ is the indicator function and $N_{\rm pair}$ is the number of matched epoch pairs. Thus $f_{\rm same}$ measures the fraction of two-band epoch pairs for which the two bands lie on the same side of their respective band-specific mean magnitudes. Unlike $J_{\rm time}$ and $J_{\phi}$, this statistic uses only the sign of the deviations and is therefore less sensitive to a small number of high-significance points.

These three statistics are used as auxiliary diagnostics for rejecting sources whose \texttt{F814W} measurements are inconsistent with the \texttt{F555W} period solution. Among sources already passing the \texttt{F555W} selection, we retain those satisfying
\begin{equation}
    \left(J_{\phi}\geq 0.5,\ 
    J_{\rm time}\geq 0.5,\ 
    f_{\rm same}\geq 0.6\right)
    \quad \mathrm{or} \quad
    f_{\rm same}\geq 0.9 .
\end{equation}
The second condition acts as a recovery criterion for sources with very high directional agreement. Because $f_{\rm same}$ counts the fraction of same-sign two-band epoch pairs but does not weight the strength of the deviations, $f_{\rm same}\geq0.9$ means that almost all paired epochs vary in the same direction. Such behavior is unlikely to arise from random noise, even when $J_{\rm time}$ or $J_{\phi}$ is modest because the \texttt{F814W} response is low-amplitude or noisier.
This two-band consistency requirement leaves 754 candidates in Field~F1 and 745 candidates in Field~F2.

We then perform a final vetting step for the sources that pass the automated screening. This step combines visual inspection of the folded two-band light curves with a long-period reanalysis when needed. The visual inspection primarily removes sources with poor light-curve quality, poorly constrained Fourier morphology, or residual two-band inconsistencies that are not fully captured by the numerical statistics. For candidates with preferred periods approaching the effective baseline of the primary data ($P \gtrsim 25$~days), we add the supplementary F555W epochs associated with the same \texttt{global\_id} and repeat the LS and model-selection procedure. We reject sources whose LS periodogram remains non-convergent, i.e., whose highest-power structure is a broad ridge rather than a localized peak. For one source, \texttt{F2-26116}, visual inspection of both bands favors the 7.33445~d solution over the automated 3.24590~d solution. This final vetting removes 34 sources in Field~F1 and 48 in Field~F2, leaving 720 and 697 candidates, respectively. The final catalog therefore contains 1417 variable-star candidates across Fields~F1 and~F2. In Section~5, we define a secure classical Cepheid subset and characterize the remaining variables at the candidate-population level rather than enforcing complete physical classifications.

\section{Sample Refinement and Variable-Star Population Analysis}

\subsection{Sample Refinement and Cepheid Classification}

After the variable-selection stage, we perform source-by-source local artificial-star tests (ASTs) to quantify crowding/blending biases and to estimate realistic local photometric uncertainties. For each candidate, we inject 100 single artificial stars at random positions within 2.4\arcsec\ of the target, using the pre-correction mean magnitude as the input brightness \citep{2022ApJ...934L...7R}. Injecting one artificial star at a time avoids artificially increasing the local crowding. We then rerun the same photometric pipeline and quality filtering used for the real data. From the input--output magnitude-difference distribution, we estimate the local blending bias and the corrected uncertainty. These bias corrections are applied only to the mean magnitudes used for classification and derived color/Wesenheit quantities, not to the individual epoch measurements.

\begin{figure*}
    \centering
    \includegraphics[width=\linewidth]{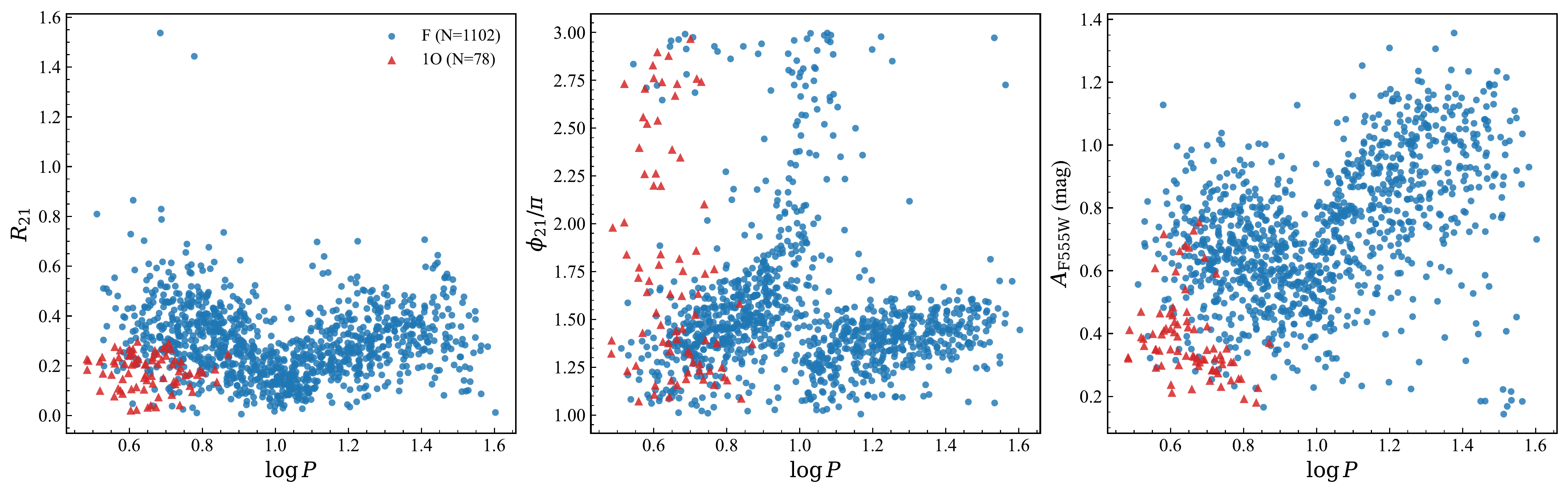}
    \caption{Fourier-parameter distributions of the secure Cepheid sample, including $\log P$ versus $R_{21}$, $\phi_{21}$, and the \texttt{F555W} amplitude. Fundamental-mode (F) Cepheids are shown as blue circles, and first-overtone (1O) Cepheids are shown as red triangles.}
    \label{fig:fourier_params}
\end{figure*}

We prioritize classical Cepheids because they generally have higher-quality photometry and are strongly constrained by characteristic colors and period--luminosity/Wesenheit relations. The refinement workflow is:
\begin{enumerate}
\item Color pre-selection. We require $(\mathrm{F555W}-\mathrm{F814W})\in[0.3,2.0]$. The red limit is intentionally permissive to retain potentially reddened Cepheids. This step leaves 1317 candidates.

\item Random-forest mode classification (F vs 1O). We train a random-forest model on OGLE LMC Cepheids \citep{2015AcA....65..297S,2017AcA....67..103S} to separate fundamental-mode (F) and first-overtone (1O) pulsators. The input features are $\log P$, peak-to-peak amplitude $A$, Fourier amplitude ratio $R_{21}$, and the Wesenheit residual $\Delta W_{VI}$ relative to an F-mode ridge line. Light-curve morphology is measured from the second-order Fourier fit in \texttt{F555W}. Fourier parameters such as $R_{21}$ and $\phi_{21}$ provide mode-sensitive information on Cepheid light-curve shape and have been widely used in OGLE Cepheid analyses \citep[e.g.,][]{1981ApJ...248..291S,1986A&A...169..149A,2009A&A...507.1729D,2015AcA....65..297S,2017AcA....67..103S}. In the present classifier, we use $R_{21}$ as the Fourier-shape input and retain $\phi_{21}$ as an additional cataloged morphology diagnostic. Supervised classifiers can combine morphology, period, amplitude, color, and luminosity information rather than relying on a single diagnostic \citep[e.g.,][]{2011MNRAS.414.2602D,2018MNRAS.477.3145J,2019A&A...625A..97R}. We define
\[
W_{VI}=I-1.55(V-I),
\]
\[
\Delta W_{VI}=W_{VI}-\left[a(\log P-1)+b\right].
\]
To avoid information leakage, the reference F-mode ridge is fitted only on the training F-mode subset using iterative $3\sigma$ clipping, with the slope fixed to $-3.3$:
\[
W_{VI}=-3.3(\log P-1)+12.5800.
\]
With an 80\%/20\% stratified train/test split, the held-out balanced accuracy is 0.9931. Feature importance ranks $\Delta W_{VI}$ first, followed by $R_{21}$ and amplitude, while the contribution from $\log P$ is secondary. In the homogeneous training sample, this internal accuracy indicates that the adopted feature set captures the expected separation between F and 1O Cepheids.

When applying the classifier to M101, we construct the same-form features while avoiding an external distance prior. For each color-selected candidate, we compute $W_{VI}=\mathrm{F814W}-1.55(\mathrm{F555W}-\mathrm{F814W})$ from the AST-corrected mean magnitudes. We use the same fixed slope, $-3.3$, and determine the intercept from the subset of our sources that are cross-matched to the Cepheid catalog of \citet{2011ApJ...733..124S}. The external catalog is used only to define this high-purity reference subset. The fitted intercept is based on our AST-corrected magnitudes and adopted periods, rather than on an external distance modulus or external photometry. The resulting reference ridge provides the $\Delta W_{VI}$ feature for the random-forest classifier. All color-selected sources remain available for the subsequent HST-system PW refinement.

In an ideal high-S/N and low-crowding data set, F and 1O Cepheids should occupy relatively distinct PL/PW ridges. In M101, unresolved companions, compact associations, local background, crowding, and lower signal-to-noise broaden these loci and make mode assignment based only on PW residuals fragile \citep[e.g.,][]{2018ApJ...861...36A,2021ApJ...910..118P,2022ApJ...934L...7R}. The RF classifier therefore combines $\Delta W_{VI}$ with period, amplitude, and the Fourier morphology parameter $R_{21}$, while $\phi_{21}$ is retained as an additional catalog diagnostic and shown in Figure~\ref{fig:fourier_params}. This mode-aware treatment follows the well-established F/1O separation in the Milky Way and LMC \citep{2020ApJS..249...18C} and reduces the risk that unrecognized 1O Cepheids broaden the final PW relation.

\item PW-based contaminant rejection after mode correction. For objects predicted as 1O, we first convert to equivalent fundamental-mode periods using
\[
P_{0}=\frac{P_{1}}{0.716-0.027\log P_{1}},
\]
following Petersen-relation-based calibrations \citep{1997MNRAS.286L...1F,2008A&A...488..731R}. We then construct the HST-system Wesenheit magnitude
\[
W_{\mathrm{H}}=\mathrm{F814W}-1.3(\mathrm{F555W}-\mathrm{F814W}),
\]
\citep{2019ApJ...876...85R,2022ApJ...940...64Y} and iteratively clip under
\[
\left|W_{\mathrm{H},i}-W_{\mathrm{H,fit}}\right|\le 3\sigma,
\]
with fixed slope
\[
\frac{\mathrm{d}W_{\mathrm{H,fit}}}{\mathrm{d}\log P}=-3.3,
\]
to reject remaining contaminants and define a compact, self-consistent secure Cepheid sequence in the HST system. The final secure sample contains 1102 F-mode and 78 1O-mode classical Cepheids. Their parameter distributions are shown in Figure~\ref{fig:fourier_params}. Sources outside this secure subset are retained in the broader variable-candidate pool.
\end{enumerate}

\begin{figure*}
    \centering
    \includegraphics[width=\linewidth]{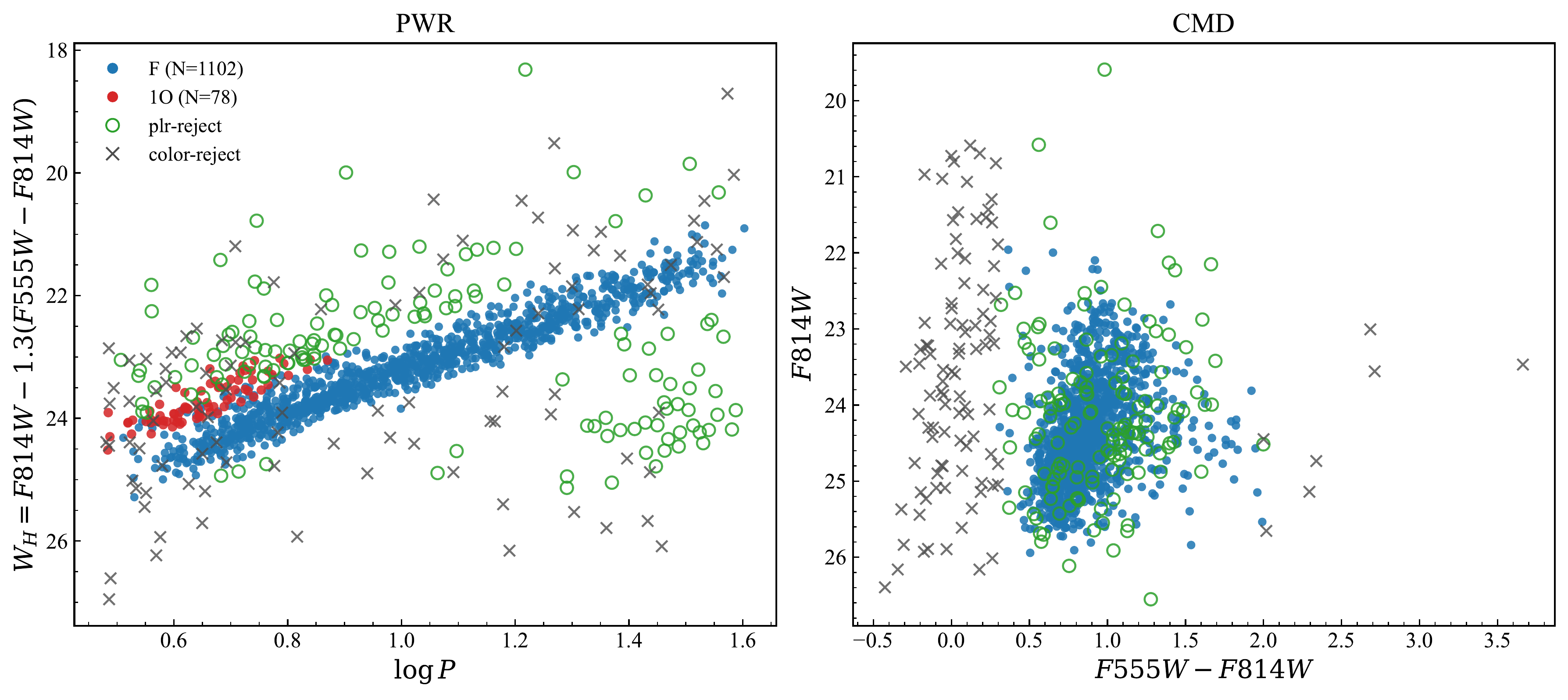}
    \caption{Selection diagnostics in the period--Wesenheit and color--magnitude planes. \emph{Left}: period--Wesenheit relation (PWR). Filled circles mark the final secure Cepheid sample, with fundamental-mode (F) Cepheids shown in blue and first-overtone (1O) Cepheids shown in red. Green open circles denote sources that satisfy the color pre-selection but are rejected by the iterative PWR clipping, while gray crosses denote sources rejected by the color criterion. The secure Cepheids form a narrow PL/PW sequence, whereas the non-secure variable candidates occupy a much broader magnitude range. This contrast is used in Section~\ref{sec:comparison_previous} as an empirical check that the newly identified Cepheids are not dominated by randomly distributed false positives. \emph{Right}: color--magnitude diagram (CMD). Secure Cepheids are shown as blue filled circles (without separating F and 1O), and the other two rejected groups use the same symbols as in the left panel.}
    \label{fig:pwr_cmd}
\end{figure*}

The F and 1O samples occupy clearly separated regions in the Fourier-parameter diagrams and in the PW plane (Figures~\ref{fig:fourier_params} and~\ref{fig:pwr_cmd}), consistent with the behavior observed for Galactic and LMC Cepheids. This separation supports the mode-assignment procedure and the use of light-curve morphology when constructing Cepheid samples in crowded extragalactic fields. It also supports treating F and 1O Cepheids separately in nearby-host HST samples, where this distinction is often absent from distance-ladder-oriented catalogs.

Because the random-forest model is designed only to separate F and 1O Cepheids, non-Cepheid identification is driven mainly by the PW-based refinement rather than by the classifier itself. The 1O training set is concentrated at short periods ($\lesssim 7$~days), so longer-period objects near the 1O locus can be preferentially assigned to F, bypassing fundamentalization and increasing PLR scatter. Residual contamination can therefore remain outside the secure Cepheid subset. Sources above the Cepheid relation are likely overluminous Cepheids, such as binaries or cluster-blended systems \citep{2018ApJ...861...36A}. Short-period sources below the relation may include anomalous Cepheids (ACEPs) \citep{2008AcA....58..293S}. A small fraction of long-period objects ($P>20$~days) may be exceptionally luminous T2Cs \citep{2026ApJ...997L..24C}. The resulting distributions in PW and CMD space are shown in Figure~\ref{fig:pwr_cmd}.

\begin{figure*}
    \centering
    \includegraphics[width=\linewidth,height=0.80\textheight,keepaspectratio]{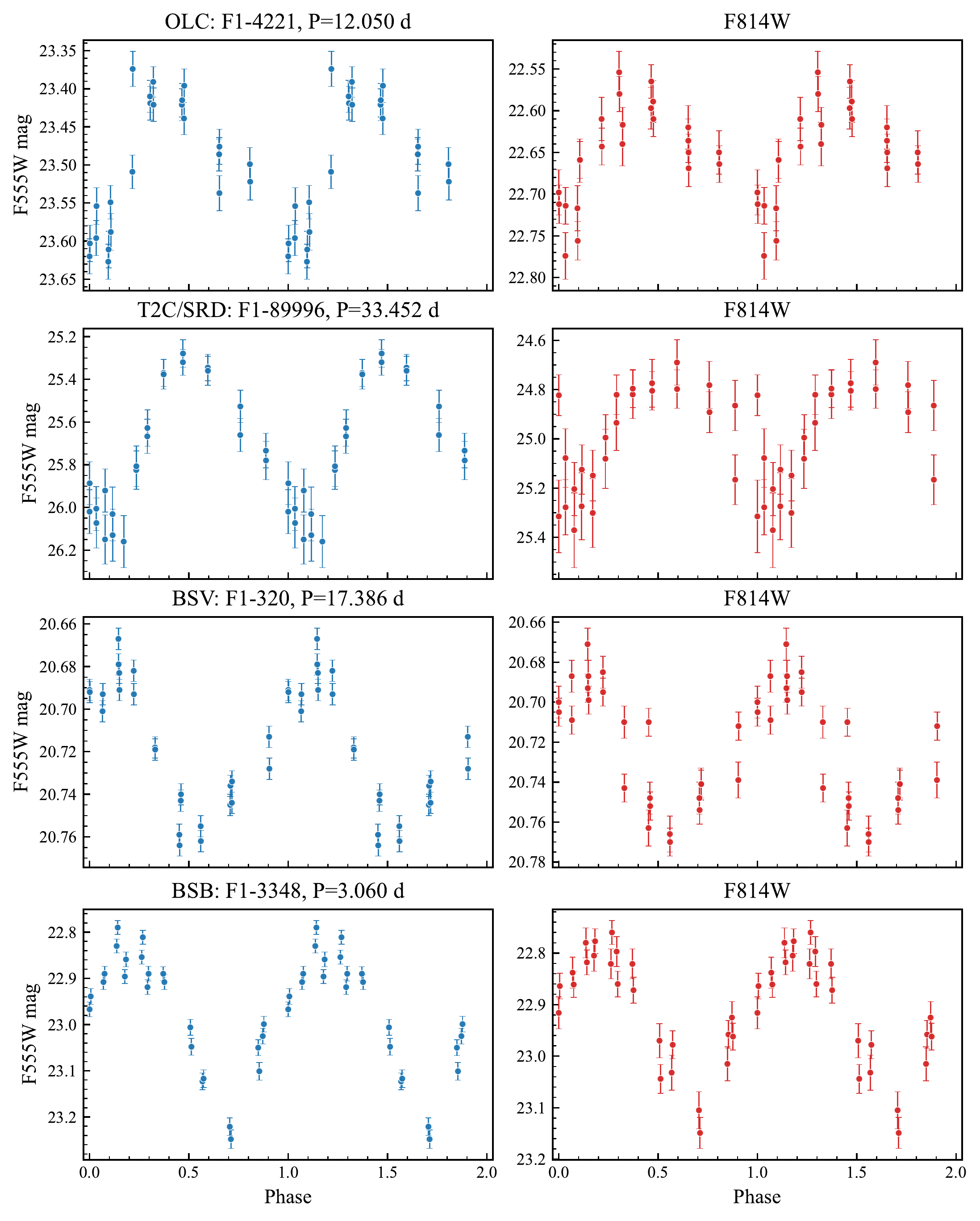}
    \caption{Representative two-band phase-folded light curves for four candidate groups. From top to bottom: overluminous Cepheid candidates, Type~II Cepheid/yellow semi-regular variable candidates, blue-supergiant variable candidates, and blue-supergiant binary candidates. The displayed examples are \texttt{F1-4221}, \texttt{F1-89996}, \texttt{F1-320}, and \texttt{F1-3348}, respectively. The left column shows \texttt{F555W}, and the right column shows \texttt{F814W}.}
    \label{fig:four_class_folded}
\end{figure*}

\subsection{Possible Other Variable-Star Populations}

Outside the secure classical Cepheid subset, we examine objects rejected by the color and/or PW criteria. These rejection paths are methodological rather than physically exclusive: different populations can overlap in period, luminosity, color, amplitude, and light-curve morphology. We therefore treat the following labels as provisional candidate classes for descriptive and population-level discussion rather than as definitive source-by-source classifications.

A strong concentration of rejected sources lies above the classical Cepheid PW relation, mostly within $\sim$1.5~mag of the ridge. Most of these sources still satisfy the Cepheid color cut, while a smaller fraction fails it. This pattern is consistent with overluminous or environmentally brightened Cepheids rather than a wholly distinct class \citep{2021ApJ...910..118P}. Unresolved companions, compact associations, or local crowding can increase the mean flux, shifting objects upward in PW space, and can also modify the integrated color. Many such objects show reduced observed amplitudes, as expected from amplitude dilution by additional non-variable light.

Sources below the Cepheid sequence are more plausibly intrinsically fainter pulsators. Their positions in period--luminosity space and overall photometric behavior suggest that most belong to the T2C/yellow semi-regular regime \citep{2017A&A...603A..70G}, with a minority of short-period objects potentially overlapping the ACEP range. Given the uncertain empirical boundaries among these classes in crowded extragalactic fields, we retain them as candidate groups.

In the bright blue region, rejected sources separate empirically into two groups by amplitude and morphology: (i) low-amplitude objects (typically $\lesssim 0.2$~mag, sometimes $<0.1$~mag), consistent with blue-supergiant-like variability, and (ii) higher-amplitude objects (up to $\sim 0.5$~mag) with distinct phased morphologies, conservatively labeled as blue-supergiant binary candidates. Together with the overluminous Cepheid candidates and the short-period faint-side T2C/SRD candidates discussed above, these empirical groupings contain 96 overluminous Cepheid candidates, 94 T2C/SRD candidates, 24 blue-variable candidates, and 23 blue-supergiant-binary candidates. Representative two-band phased light curves for these four candidate groups are shown in Figure~\ref{fig:four_class_folded}. These examples show empirical differences in location, amplitude, and morphology; they are not intended as physical templates, and the labels remain provisional candidate-level classifications.

\subsection{Catalog Description}

We release the final M101 variable-star census as a machine-readable catalog. For each source, the catalog provides the field and source identifier, sky position, sampling information, period-search diagnostics, light-curve parameters from the adopted \texttt{F555W} fit, two-band consistency statistics, AST-corrected mean photometry, derived color/Wesenheit quantities, and the final candidate class.

Principal columns include \texttt{field}, \texttt{global\_id}, coordinates (\texttt{ra}, \texttt{dec}), sampling counts (\texttt{n\_f555w}, \texttt{n\_f814w}), the adopted period (\texttt{best\_period}), period-search quantities (\texttt{fap}, \texttt{max\_power}, \texttt{period\_snr}, \texttt{time\_span}), \texttt{F555W} fit parameters (\texttt{total\_amplitude\_f555w}, \texttt{r21}, \texttt{phi21}, \texttt{rmse\_f555w}, \texttt{r\_squared\_f555w}, \texttt{dbic\_f555w}), two-band statistics (\texttt{N\_pair}, \texttt{f\_same}, \texttt{J\_time}, \texttt{J\_phi}), AST-corrected mean magnitudes (\texttt{mean\_mag\_f555w\_astcorr}, \texttt{mag\_f814w\_astcorr}), derived \texttt{color\_VI}, \texttt{W}, \texttt{logP}, and the final fields \texttt{is\_cepheid\_secure}, \texttt{class}, and \texttt{classification\_note}. Additional columns retain DOLPHOT quality diagnostics and AST uncertainty terms.

Besides the summary catalog, we also release source-level two-band light curves for every object in \texttt{F555W} and \texttt{F814W}. Table~\ref{tab:m101_catalog} shows a representative subset of 30 entries from the machine-readable catalog. Class abbreviations used in this work are BSV = blue-supergiant variable candidate, BSB = blue-supergiant binary candidate, T2C/SRD = Type~II Cepheid/yellow semi-regular candidate, and OLC = overluminous Cepheid candidate.

\onecolumngrid
\setlength{\LTleft}{0pt}
\setlength{\LTright}{0pt}
\setlength{\tabcolsep}{3pt}
\tiny
\begin{longtable}{lrrrrrrrrrl}
\caption{M101 variable-star catalog.}\label{tab:m101_catalog} \\
\hline
\texttt{global\_id} & \texttt{ra} & \texttt{dec} & $P$ & $\langle F555W\rangle_c$ & $\langle F814W\rangle_c$ & $A_{555}$ & $R_{21}$ & $\phi_{21}$ & $R^2_{555}$ & \texttt{class} \\
\hline
\endfirsthead

\hline
\texttt{global\_id} & \texttt{ra} & \texttt{dec} & $P$ & $\langle F555W\rangle_c$ & $\langle F814W\rangle_c$ & $A_{555}$ & $R_{21}$ & $\phi_{21}$ & $R^2_{555}$ & \texttt{class} \\
\hline
\endhead

\hline
\endfoot

\hline
\multicolumn{11}{l}{\textit{Note.} This table shows the first 30 entries. The full catalog is available in machine-readable form.} \\
\endlastfoot

F1-100123 & 210.890295 & 54.326409 & 4.872 & 25.822 & 25.175 & 0.694 & 0.264 & 1.244 & 0.908 & F \\
F1-100282 & 210.919527 & 54.376617 & 5.196 & 25.820 & 25.063 & 0.758 & 0.407 & 1.355 & 0.919 & F \\
F1-101094 & 210.900546 & 54.331205 & 5.567 & 25.862 & 25.019 & 0.571 & 0.227 & 1.624 & 0.778 & F \\
F1-10114 & 210.873406 & 54.378824 & 25.794 & 24.194 & 23.324 & 1.106 & 0.597 & 1.486 & 0.839 & F \\
F1-101376 & 210.856659 & 54.390565 & 5.633 & 25.914 & 25.137 & 0.843 & 0.430 & 1.609 & 0.907 & F \\
F1-102500 & 210.933785 & 54.370217 & 4.852 & 25.931 & 25.166 & 0.786 & 0.532 & 1.431 & 0.931 & F \\
F1-1026 & 210.868212 & 54.361946 & 5.565 & 22.239 & 21.605 & 0.198 & 0.255 & 2.103 & 0.855 & OLC \\
F1-102643 & 210.860174 & 54.331950 & 5.172 & 25.720 & 25.349 & 0.815 & 0.208 & 1.323 & 0.689 & T2C/SRD \\
F1-10302 & 210.863153 & 54.360680 & 24.812 & 24.477 & 23.312 & 1.011 & 0.390 & 1.461 & 0.886 & F \\
F1-10368 & 210.862752 & 54.356109 & 22.535 & 25.069 & 23.579 & 0.857 & 0.412 & 1.488 & 0.941 & F \\
F1-10452 & 210.862102 & 54.344594 & 24.805 & 24.525 & 23.306 & 1.000 & 0.363 & 1.669 & 0.865 & F \\
F1-10476 & 210.879717 & 54.333520 & 16.978 & 24.202 & 23.395 & 0.955 & 0.205 & 1.410 & 0.946 & F \\
F1-105351 & 210.858525 & 54.334385 & 6.031 & 25.956 & 25.281 & 0.661 & 0.257 & 1.694 & 0.803 & F \\
F1-10544 & 210.857435 & 54.388809 & 20.185 & 24.333 & 23.422 & 1.062 & 0.245 & 1.379 & 0.972 & F \\
F1-10652 & 210.870938 & 54.357203 & 19.984 & 24.431 & 23.338 & 0.581 & 0.241 & 1.311 & 0.967 & F \\
F1-10718 & 210.832360 & 54.360212 & 4.358 & 23.939 & 23.885 & 0.271 & 0.526 & 1.736 & 0.838 & BSB \\
F1-107877 & 210.909660 & 54.345338 & 5.611 & 25.920 & 25.078 & 0.775 & 0.445 & 1.531 & 0.807 & F \\
F1-107923 & 210.928088 & 54.365664 & 5.191 & 25.799 & 25.162 & 0.579 & 0.548 & 1.412 & 0.853 & F \\
F1-108014 & 210.912580 & 54.347290 & 5.014 & 25.822 & 25.142 & 0.700 & 0.259 & 1.354 & 0.906 & F \\
F1-10832 & 210.865336 & 54.369235 & 20.899 & 24.494 & 23.419 & 1.160 & 0.460 & 1.637 & 0.920 & F \\
F1-10937 & 210.928893 & 54.367331 & 17.882 & 24.176 & 23.447 & 0.757 & 0.390 & 1.386 & 0.898 & F \\
F1-109677 & 210.879881 & 54.328320 & 5.684 & 25.872 & 25.103 & 0.910 & 0.655 & 1.250 & 0.868 & F \\
F1-109791 & 210.912386 & 54.346767 & 4.410 & 25.821 & 25.199 & 0.798 & 0.244 & 1.438 & 0.913 & F \\
F1-10980 & 210.857997 & 54.344082 & 23.451 & 24.485 & 23.361 & 1.022 & 0.301 & 1.498 & 0.973 & F \\
F1-10994 & 210.860046 & 54.366078 & 9.518 & 24.745 & 23.241 & 0.507 & 0.108 & 1.272 & 0.835 & OLC \\
F1-11050 & 210.856119 & 54.352767 & 15.961 & 24.214 & 23.447 & 0.680 & 0.247 & 1.402 & 0.966 & F \\
F1-11068 & 210.850116 & 54.378963 & 19.102 & 24.273 & 23.392 & 0.309 & 0.412 & 1.093 & 0.827 & F \\
F1-111683 & 210.933996 & 54.371564 & 4.382 & 25.918 & 25.119 & 0.576 & 0.437 & 1.084 & 0.857 & F \\
F1-112886 & 210.930681 & 54.369528 & 5.046 & 25.853 & 25.165 & 0.637 & 0.458 & 1.345 & 0.893 & F \\
F1-11399 & 210.882397 & 54.373581 & 14.345 & 23.952 & 23.992 & 0.158 & 0.528 & 1.109 & 0.753 & T2C/SRD \\
\end{longtable}
\setlength{\tabcolsep}{6pt}
\normalsize
\twocolumngrid

\section{Discussion}

We compare the catalog with previous Cepheid samples and then discuss observing-design limitations and the broader variable-star content of the time series.

\subsection{Comparison with previous Cepheid samples}
\label{sec:comparison_previous}

We first compare our catalog with the widely used Cepheid catalog of \citet{2011ApJ...733..124S}. Their catalog contains 827 sources in the region considered here. We solve a field-specific astrometric shift from high-confidence counterparts and then perform a one-to-one source assignment using positional separation and two-band photometric consistency. This procedure identifies 778 counterparts in our variable catalog, while 49 literature sources do not have final-catalog counterparts. Of the matched sources, 760 are retained in our secure F/1O classical Cepheid sample. The remaining 18 are retained in the broader variable catalog but are not classified as secure classical Cepheids.

\begin{table*}[t]
\centering
\caption{Cross-match with the Shappee \& Stanek Cepheid catalog.}
\label{tab:shappee_match}
\begin{tabular}{lrrrrl}
\hline
\hline
Period range & Total & Matched & Match rate & Non-secure Cep. & Non-secure classes \\
\hline
$P < 10$ d          & 378 & 350 & 92.6\% & 10 & 8 OLC, 2 T2C/SRD \\
$10 \leq P < 20$ d & 306 & 306 & 100.0\% & 7  & 7 OLC \\
$20 \leq P < 30$ d & 89  & 85  & 95.5\% & 0  & -- \\
$30 \leq P < 40$ d & 35  & 30  & 85.7\% & 0  & -- \\
$P \geq 40$ d      & 19  & 7   & 36.8\% & 1  & 1 OLC \\
\hline
Total               & 827 & 778 & 94.1\% & 18 & 16 OLC, 2 T2C/SRD \\
\hline
\end{tabular}
\tablecomments{``Non-secure Cep.'' refers to cross-matched sources that are present in our variable catalog but are not retained in the final secure F/1O classical Cepheid sample.}
\end{table*}

\begin{figure*}[t]
\centering
\includegraphics[width=0.63\textwidth]{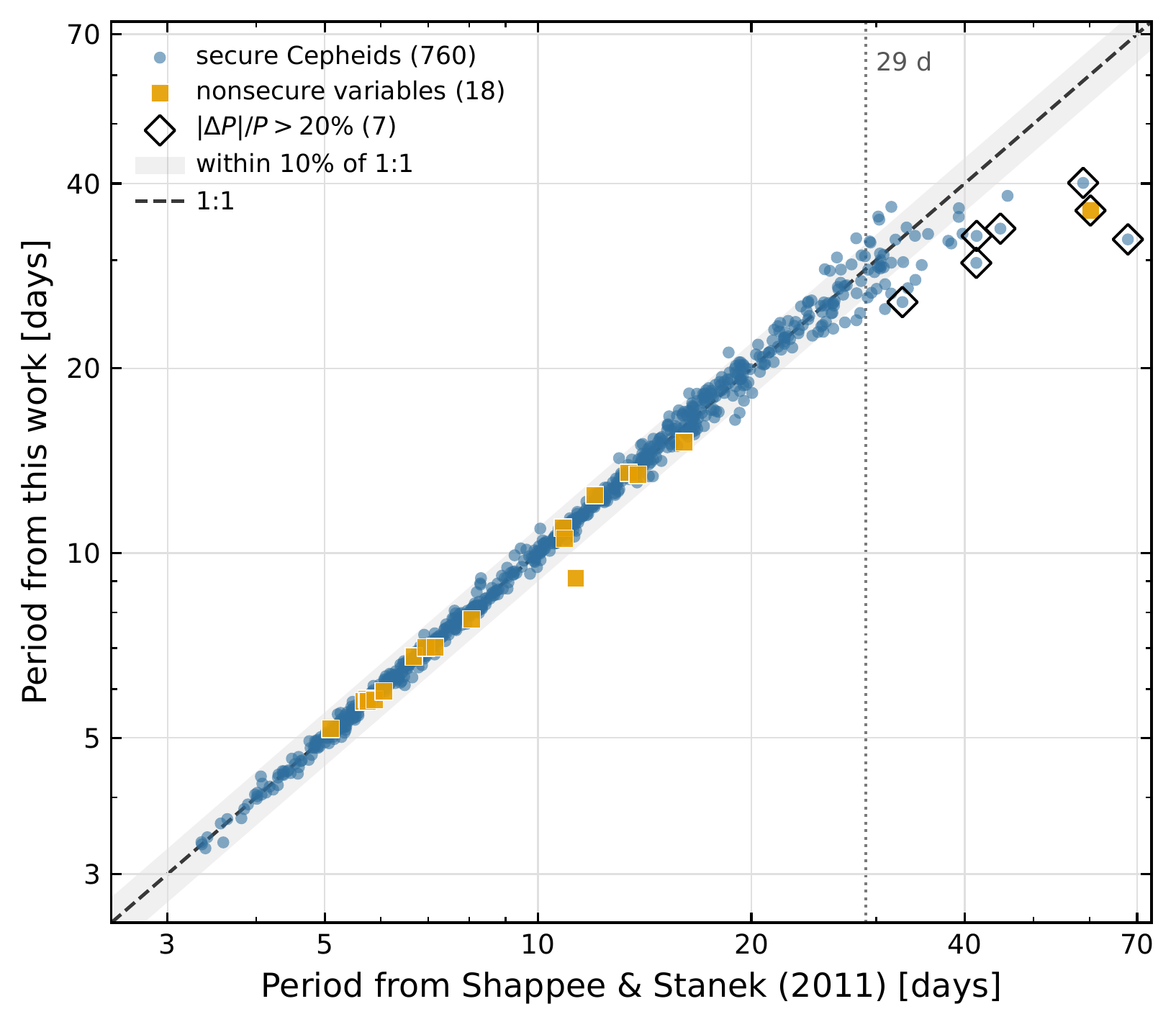}
\caption{Period comparison for the 778 cross-matched sources between this work and \citet{2011ApJ...733..124S}. Blue circles mark the 760 counterparts retained as secure Cepheids, orange squares mark the 18 matched nonsecure variables, and black open diamonds mark the seven sources with period differences larger than 20\%. The gray band encloses periods within 10\% of one-to-one agreement. The dashed line marks one-to-one agreement, and the vertical dotted line indicates the approximate 29~d baseline of the primary ACS data.}
\label{fig:shappee_period_comparison}
\end{figure*}

Table~\ref{tab:shappee_match} summarizes the comparison as a function of the Shappee \& Stanek period. The overlap is high at $P<30$~d, where the match fraction is 92.6--100.0\%, but it drops to 36.8\% at $P\geq40$~d. This behavior is consistent with the limited $\sim$29~d baseline of the primary ACS time series, where period recovery becomes more sensitive to phase coverage and alias structure. For candidates with $P>25$~d, the supplementary \texttt{F555W} epochs generally produce modest period refinements, with most revised periods differing by less than 10\% from the primary-data solutions. Only two sources show large period changes, where the archival epochs fill previously missing peak or trough phases. The sparse supplementary data therefore mainly provide a stability check for selected long-period candidates rather than systematically reshaping the sample.

Figure~\ref{fig:shappee_period_comparison} shows that the cross-matched periods generally follow the one-to-one relation, with the largest discrepancies at the long-period end. For the conspicuous offset near 10~d, the source has two comparable Lomb--Scargle peaks at 9.1 and 11.1~d. We select the 9.1~d peak, whereas the Shappee \& Stanek period of 11.3~d corresponds to the other peak. The 18 matched sources not retained as secure classical Cepheids comprise 16 overluminous Cepheid candidates and two T2C/SRD candidates. For sources recovered by both searches, the remaining differences mainly reflect long-period ambiguity and final class assignment.

The comparison also shows that our catalog contains 420 newly identified secure F/1O Cepheids beyond the Shappee \& Stanek sample. To test whether this sample increase depends on sources admitted by the relaxed initial S/N or crowding cuts, we apply stricter PHANGS-like post-selection cuts of ${\rm S/N}>5$ and crowding $\leq0.67$ in both bands. These cuts retain 1381 of the 1417 variables (97.5\%) and 1156 of the 1180 secure classical Cepheids (98.0\%). Table~\ref{tab:photometric_quality_robustness} separates the secure sample by catalog provenance. The stricter selection retains 758 of the 760 secure Cepheids matched to \citet{2011ApJ...733..124S} (99.7\%) and 398 of the 420 newly identified secure Cepheids (94.8\%). The sample increase is therefore not primarily driven by the more permissive entrance-level photometric-quality cuts, although this test does not imply complete insensitivity to photometric quality.

\begin{table}[t]
\centering
\caption{Sensitivity of the Cepheid sample to stricter photometric-quality cuts.}
\label{tab:photometric_quality_robustness}
\begingroup
\small
\renewcommand{\arraystretch}{1.05}
\begin{tabular*}{\columnwidth}{@{\extracolsep{\fill}}lrrr@{}}
\hline\hline
Sample & Fiducial & Stricter & Survival \\
\hline
All secure   & 1180 & 1156 & 98.0\% \\
S\&S matched &  760 &  758 & 99.7\% \\
New          &  420 &  398 & 94.8\% \\
\hline
\end{tabular*}
\vspace{3pt}
\parbox{\columnwidth}{\footnotesize\raggedright\textit{Note.} The stricter cuts adapt the PHANGS-HST cluster-candidate criteria of \citet{2022MNRAS.509.4094T}: ${\rm S/N}>5$ and crowding $\leq0.67$, required here in both bands. ``S\&S matched'' and ``New'' partition the secure sample.}
\endgroup
\end{table}

In Figure~\ref{fig:pwr_cmd}, the 420 newly identified secure Cepheids lie on a narrow Cepheid PL/PW sequence with a width of about 1~mag, whereas 237 other variable candidates occupy a much broader range of about 6~mag. If random false positives were drawn from this broad distribution, only about $(1/6)\times237$, or roughly 40, would be expected to fall into the Cepheid sequence. Conversely, if all 420 newly identified Cepheids were false positives from the same broad distribution, one would expect more than 2500 false variables across the full 6~mag range, which is not observed. This simple width-based comparison is intended only as a qualitative consistency check, rather than a precise quantitative contamination estimate, because the background need not be uniform in PW space.

Source-finding completeness at the catalog-entrance stage is one contributor to these newly identified Cepheids. As discussed in Section~3, our multi-epoch DOLPHOT catalog is built on a common astrometric frame and source list, whereas a two-band single-epoch reference-catalog entrance can only extract light curves for sources detected in the reference images and matched between filters. The catalog-entrance difference quantified above and the global AST curves in Figure~\ref{fig:ast_m101f2} indicate losses at the level of a few hundred relevant sources before variability information is used. Specifically, estimating the two-band entrance probability for each secure Cepheid as $C_{555}(m_{555})C_{814}(m_{814})$ gives a mean two-band completeness of 0.82, corresponding to more than 200 expected losses from two-band photometric completeness alone. This scale can account for a substantial fraction of the newly identified secure Cepheids.

Another contributor is 1O recovery: 63 of the 78 secure 1O Cepheids are newly identified relative to the Shappee \& Stanek catalog. Template-driven searches that do not explicitly separate F and 1O pulsators can be less efficient for nearly sinusoidal, lower-amplitude 1O light curves.

We also note that the Cepheid catalog used by \citet{2022ApJ...934L...7R} was constructed for the WFC3 region used in their distance-ladder analysis, whereas our catalog is based on the ACS fields analyzed here. The sky coverage is therefore not identical. Their reported M101 Cepheid sample extends to a maximum period of 29.5~d, close to the baseline of the primary ACS time series.

\subsection{Implications for future time-domain studies}

The comparison also points to two limitations of the observing design. First, although the M101 series is two-band, the F555W and F814W cadences are very similar, so their window functions are effectively the same. Two-band agreement therefore confirms coherent color behavior, but it does not provide a fully independent test of period uniqueness, because aliases or window-function ambiguities in one band tend to appear in the other. Future extragalactic time-domain programs would benefit from intentionally offset multi-band cadences that provide complementary temporal constraints.

Second, the catalog shows that HST time-series data in nearby SN-host galaxies contain broader variability information than is usually exploited in distance-ladder analyses. Additional luminous variable populations are already detectable outside the secure Cepheid locus, but their classification is limited by cadence design, epoch number, and less mature empirical frameworks for non-Cepheid variables in crowded extragalactic fields. Even modest cadence expansion could improve period recovery, classification confidence, and population-level interpretation for variables brighter than most Cepheids.

\section{Conclusions}

We present a homogeneous HST/ACS census of variable stars in two disk fields of M101 based on the primary F555W/F814W time-series data, with a small number of archival F555W epochs used only to test and refine selected long-period solutions. The final catalog contains 1417 variable-star candidates. From this parent sample, we define a secure classical Cepheid subset comprising 1102 fundamental-mode and 78 first-overtone Cepheids.

The final sample is not based on a single empirical variability cut. The workflow combines Lomb--Scargle period searches, Fourier light-curve modeling, two-band coherence statistics, local artificial-star-test corrections, F/1O mode assignment, and PW-based refinement. By separating variable detection, two-band confirmation, and Cepheid-sample refinement, the procedure makes the mode-aware Cepheid sample more reproducible than a search driven mainly by a single variability statistic.

The resulting catalog is broadly consistent with previous Cepheid samples in M101 for the bulk of the population. We identify 778 cross-matched objects out of 827 sources in the widely used catalog of \citet{2011ApJ...733..124S}, with the largest recovery differences appearing at the long-period end. At the same time, our secure Cepheid sample includes 420 newly identified sources beyond the Shappee \& Stanek catalog. This asymmetry reflects different selection functions: the newly identified sources mainly trace the more inclusive multi-epoch DOLPHOT entrance catalog and explicit recovery of low-amplitude 1O Cepheids, while the reduced recovery of the longest-period literature sources reflects our cautious treatment of periods approaching or exceeding the $\sim$29~d primary baseline.

Beyond the secure Cepheid subset, the census shows that existing HST time-series data in M101 contain broader information on luminous variable populations than is usually exploited in distance-ladder work. The released catalog and source-level two-band light curves provide data for subsequent analyses of Cepheid systematics, distance-scale applications, and luminous non-Cepheid variables in nearby SN-host galaxies.

\section*{Acknowledgements} 
We thank the anonymous referee for the helpful comments. This work was supported by the National Natural Science Foundation of China (NSFC) through grants 12322306, 12373028, 12633007, 12622304, 12173047, 12233009 and 12133002. S. W. and X. C. acknowledge support from the Youth Innovation Promotion Association of the CAS (grant Nos. 2023065 and 2022055). This paper is based on observations made with the NASA/ESA \textit{Hubble Space Telescope}, obtained from the MAST data archive at the Space Telescope Science Institute (STScI), which is operated by the Association of Universities for Research in Astronomy, Inc., under NASA contract NAS 5-26555.

\software{Astropy \citep{astropy:2013,astropy:2018,astropy:2022}, Matplotlib \citep{2007CSE.....9...90H}, TOPCAT \citep{2005ASPC..347...29T}}

\bibliography{sample631}
\bibliographystyle{aasjournal}

\end{document}